\documentclass[12pt]{article}

\usepackage{graphicx}
\usepackage{amssymb}
\usepackage{amsmath}
\usepackage{bm}
\usepackage{xcolor}

\newcommand{\rf}[1]{(\ref{#1})}
\newcommand{\beq}{\begin{equation}}

\newcommand{\eeq}{\end{equation}}
\newcommand{\bea}{\begin{eqnarray}}
\newcommand{\eea}{\end{eqnarray}}

\newcommand{\e}{\mbox{e}}

\newcommand{\G}{\Gamma}
\newcommand{\lam}{\lambda}
\newcommand{\Lam}{\Lambda}

\newcommand{\om}{\omega}
\newcommand{\Om}{\Omega}
\newcommand{\del}{\delta}
\newcommand{\Del}{\Delta}

\newcommand{\kp}{\kappa}

\newcommand{\oh}{\frac{1}{2}}

\newcommand{\ra}{\rangle}
\newcommand{\la}{\langle}

\newcommand{\prt}{\partial}
\newcommand{\mi}{\!-\!}

\newcommand{\plu}{\!+\!}

\newcommand{\cD}{{\cal D}}

\newcommand{\non}{\nonumber \\}

\newcommand{\UV}{\mathrm{\scriptscriptstyle UV}}

\begin{document}

\begin{center}
\vspace{24pt}
{ \large \bf IR and UV limits of  CDT and their relations to FRG}

\vspace{24pt}

{\sl J. Ambjorn}$\,^{a,b}$,
{\sl J. Gizbert-Studnicki}$\,^{c,d}$
{\sl A. G\"{o}rlich}$\,^{c,d}$ and
{\sl D. Németh}$\,^{b}$

\vspace{10pt}

{\small

{\footnotesize

$^a$~The Niels Bohr Institute, Copenhagen University\\
Blegdamsvej 17, DK-2100 Copenhagen \O , Denmark.\\
email: ambjorn@nbi.dk

\vspace{10pt}

$^b$~Institute for Mathematics, Astrophysics and Particle Physics
(IMAPP)\\ Radboud University Nijmegen, Heyendaalseweg 135, \\
6525 AJ  Nijmegen, The Netherlands\\
email:  j.ambjorn@science.ru.nl, d.nemeth@science.ru.nl

\vspace{10pt}

$^c$~Institute of Theoretical Physics, Jagiellonian University,\\
 \L ojasiewicza 11, Krak\'ow, PL 30-348, Poland.\\
email: jakub.gizbert-studnicki@uj.edu.pl, andrzej.goerlich@uj.edu.pl.

$^d$~Mark Kac Center for Complex Systems Research, \\
 \L ojasiewicza 11, Krak\'ow, PL 30-348, Poland.
}

\vspace{18pt}

{\it Lectures presented by JA at the 64 Krakow School of Theoretical Physics, \\
Zakopane, June 9-15, 2024. \\
To appear in Acta Physica Polonia B.}

}

\end{center}

\vspace{24pt}

\begin{center}
{\bf Abstract}
\end{center}

\noindent
Causal Dynamical Triangulations (CDT) is a lattice theory of quantum gravity. 
It is shown how to identify the IR and 
the UV limits of this lattice theory with similar limits studied using the continuum,
functional renormalization group (FRG) approach. The main technical tool in this study 
will be the so-called two-point function. 
It will allow us to identify a correlation length not directly 
related to the propagation of   physical degrees of freedom.

\newpage

\section{Introduction}
\label{intro}

Four-dimensional  gravity is not perturbatively renormalizable. For many years 
it has been discussed if the theory could be defined as a unitary, non-perturbative 
quantum field theory. This putative theory could  contain other terms than the 
classical Einstein-Hilbert terms in the action and these additional terms could make 
the theory UV well defined. This has been well understood since the seminal work
of Stelle \cite{stelle} where an $R^2$ term was added to the classical GR action. Unfortunately, it was not so clear how to ensure unitarity of the corresponding quantum
theory. A more general setup is known as the asymptotic safety scenario \cite{weinberg}
where one, appealing to the Wilsonian renormalization group,
tries to understand if the UV limit of a  quantum gravity theory 
can be associated with a fixed point of the renormalization group. This fixed point 
could in principle be non-Gaussian, and starting with the seminal work of 
Reuter \cite{reuter} a lot of evidence has been accumulated supporting the existence 
of such a non-Gaussian fixed point 
(see \cite{review3}, \cite{review4} and \cite{review5} for extensive 
reviews). 
The evidence for this non-Gaussian fixed point 
comes from solving the renormalization group equations using the so-called functional
renormalization group technique (FRG), that goes back to Wetterich 
(see \cite{wetterich} for a review). However,  ``solving'' FRG also means in this context a truncation of the full equation, 
and it can be difficult to judge how reliable the results are. Thus it would be reassuring if 
one could verify the FRG results using an independent calculation and since we are 
discussing non-perturbative quantum field theory, the use of lattice quantum field 
theory is natural. In that case one often has to use Monte Carlo simulations, and they
also represent an approximation, but of a different kind than used in FRG. The purpose 
of these Lectures  is to compare the two approaches.

The rest of the Lectures is organized as follows: first we recall how one can use lattice 
field theory to test the existence of a putative non-perturbative UV fixed point. Here 
we use a $\phi^4$ theory in four dimensions as a example. Next we discuss ways in which 
quantum gravity can be formulated as a lattice field theory, namely by the use of 
so-called Dynamical Triangulations (DT) or Causal Dynamical Triangulations (CDT).
In this Section we also discuss how to introduce the concept of a diffeomorphism 
invariant correlation  length in quantum gravity, and in what way it 
implies finite size scaling in the lattice quantum gravity theories.   
We then compare the lattice results (obtained by Monte Carlo simulations) with the 
simplest results obtained using the FRG  approach. The final Section contains
a discussion of the results obtained so far. 

\section{Identifying fixed points in  \boldmath{$\phi^4$} lattice theory} 

\begin{figure}[t]
\centerline{\scalebox{0.22}{\rotatebox{0}{\includegraphics{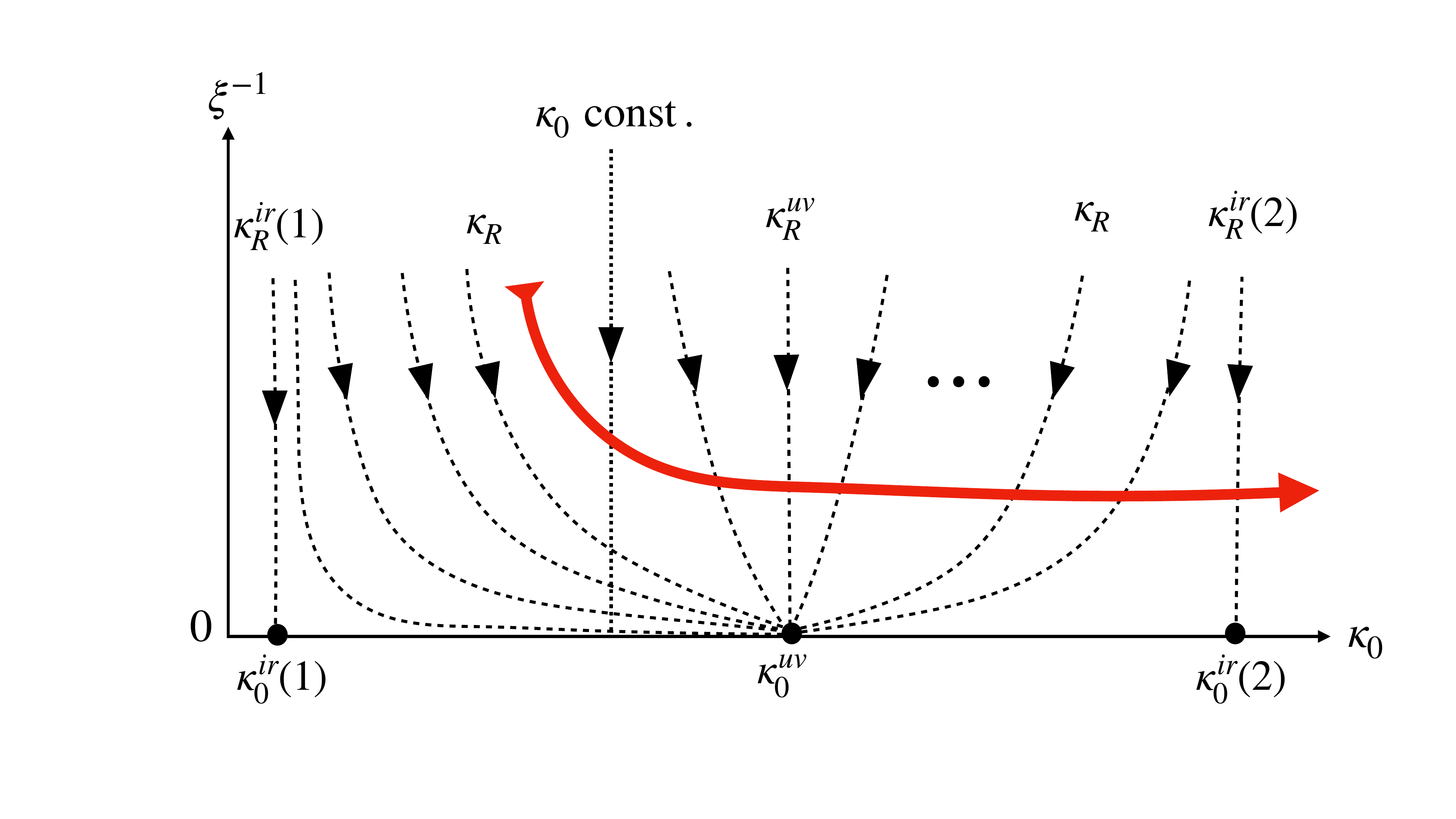}}}}
\vspace{-1cm}
\caption{ {\small The tentative $\phi^4$ phase diagram with 
an UV fixed point and two IR fixed points. The dashed lines are paths where the renormalized $\phi^4$ coupling constant $\kp_R$ is kept fixed, while on the dotted line  the bare coupling constant $\kp_0$ is fixed. The full-drawn red line illustrates the 
way the real renormalization group flow will be in a $\phi^4$ theory with a fixed $\kp_R$. It will never reach the critical line where $\xi = \infty$, and accordingly there will not be a continuum quantum field theory with a fixed $\kp_R > 0$, as first shown in \cite{luscher}.}}
\label{fig-appendix}
\end{figure}

Let us consider a $\phi^4$ lattice field theory in four dimensions. 
The lattice action used is 
\beq\label{f1}
S = \sum_n \left(  \sum_{i=1}^4 (\phi(n \plu e_i) \mi \phi(n))^2 + \mu_0 \phi^2_n + 
\kp_0 \phi^4(n) \right),
\eeq
where $n$ denotes a lattice point, where $e_i$ is a unit vector in direction $i$ and 
where the fields $\phi(n)$  and the coupling constants $\mu_0$ and $\kp_0 \geq 0$ 
are dimensionless and the length of the lattice links is 1. The theory has a 
second order phase transition line, starting at $\kp_0 =\mu_0 = 0$ and extending 
to $\kp_0 = \infty$. It separates the symmetric phase ($\la \phi \ra =0$) from the 
broken phase ($\la \phi \ra  \neq 0$). We will consider only the symmetric phase.
For each value $\mu_0,\kp_0$ one has a 
correlation length $\xi(\mu_0,\kp_0)$, defined by the exponential fall-off of the 
two-point function. It diverges when one approaches the second order 
phase transition line. Rather than using $\mu_0,\kp_0$ as variables defining
the theory, we will use $\xi,\kp_0$. Then the phase transition line will be at 
$\xi^{-1} = 0$. The possible fixed points for the theory will be on this line and 
there can potentially be both IR and UV fixed points  as shown  in Fig.\ \ref{fig-appendix}.
 For each value of the bare coupling constant $\kp_0$ we can define a 
 renormalized coupling constant $\kp_R(\kp_0,\xi)$. 
It can be expressed in terms of bare four-point and bare two-point functions 
(see \cite{munster} for details). The theory will have a UV fixed point $\kp_0^{uv}$  
if it is possible to find a path $(\xi,\kp_0(\xi))$ in the 
$\xi,\kp_0$ coupling constant plane such that
\beq\label{am1}
\kp_R(\kp_0(\xi),\xi) = \kp_R \quad {\rm for} \quad \xi \to \infty.
\eeq
Such paths for different $\kp_R$ are illustrated in Fig.\ \ref{fig-appendix}. 
Differentiating \rf{am1} wrt $\xi$ we obtain  
\beq\label{fj4}
0 = \xi\frac{d }{d \xi} \kp_R( \kp_0(\xi),\xi)=
\xi \frac{\prt \kp_R}{\prt \xi }\Big|_{\kp_0} + 
\frac{\prt \kp_R}{\prt \kp_0}\Big|_\xi\;
\xi\frac{d\kp_0}{d\xi}\Big|_{\kp_R}.
\eeq
Introducing the bare and the renormalized $\beta$-functions
\beq\label{fj5}
\beta_0(\kp_0) = \xi\frac{d\kp_0}{d\xi}\Big|_{\kp_R}, \qquad
\beta_R(\kp_R) =-\xi \frac{\prt \kp_R}{\prt \xi }\Big|_{\kp_0}
\eeq
eq.\ \rf{fj4} can be written as\footnote{The $\beta_0(\kp_0)$  function as defined 
by \rf{fj4} is strictly speaking also a function of $\xi$, but for large $\xi$ (the 
so-called scaling region) this dependence can be ignored. The same remarks are true for 
$\beta_R(\kp_R)$.} 
\beq\label{fj6}
\beta_R(\kp_R) = \frac{\prt \kp_R}{\prt \kp_0} \;\beta_0(\kp_0).
\eeq
A typical $\beta_0(\kp_0)$ is shown in Fig.\ \ref{fig-2am}, and solving \rf{fj5} for 
a fixed $\kp_R$ close to the fixed point $\kp_0^{uv}$ we obtain:
\begin{figure}[t]
\vspace{-1cm}
\centerline{\scalebox{0.2}{\rotatebox{0}{\includegraphics{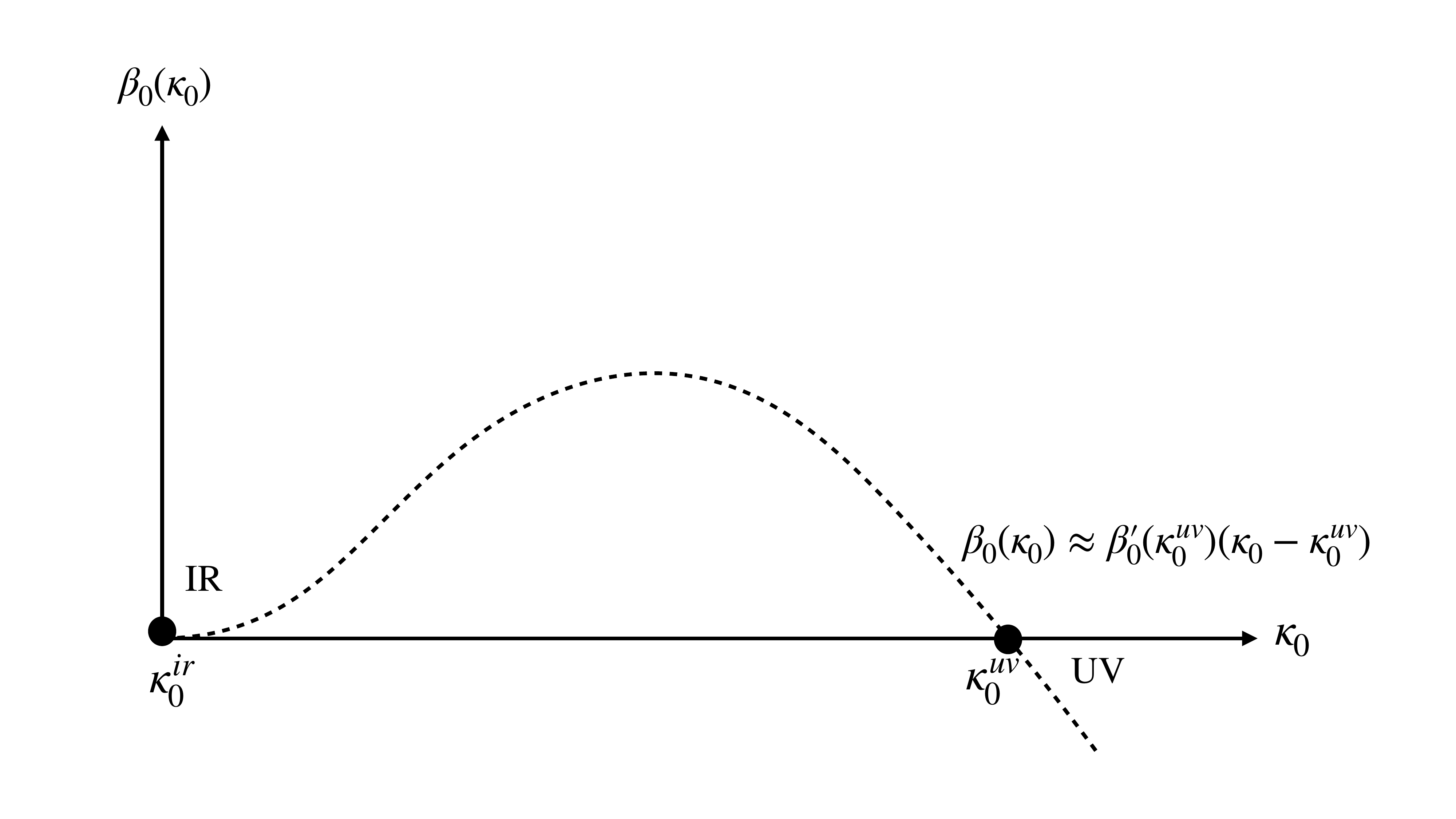}}}}
\vspace{-5mm}
\caption{\small The expected form of the $\phi^4$ $\beta$-function {\it if} the $\phi^4$ 
theory would have a UV fixed point.}
\label{fig-2am}
\end{figure}
\beq\label{am2}
 \xi\frac{d\kp_0}{d\xi}\Big|_{\kp_R} = \beta'_0(\kp_0^{uv})(\kp_0 - \kp_0^{uv}), 
 \quad {\rm i.e.} \quad 
|\kp_0(\xi)-  \kp_0^{uv}| =  c(\kp_R) \,\xi^{\beta'_0(\kp_0^{uv})},
\eeq
i.e.\ $\kp_0(\xi)  \to \kp_0^{uv}$ for $\xi \to \infty$ if $\beta_0'(\kp_0^{uv}) < 0$. 
$\kp_0^{uv}$ thus serves as an UV fixed point. 
Similarly, solving \rf{fj5} for $\kp_R$ as a function of $\xi$ for fixed 
$\kp_0$, it is seen from Fig.\ \ref{fig-appendix} that $\kp_R(\xi)$ flows to an IR fixed 
point $\kp_R^{ir}$ for $\xi \to \infty$.

In eq.\ \rf{f1} we assumed the lattice spacing was 1. One can instead  
introduce a lattice spacing of
length $a$ in \rf{f1} and $a$ will then act as an adjustable UV cut-off. At a UV fixed point 
one can define a ``continuum limit'' where $a \to 0$ (and $\kp_R > 0$) in the following 
way: introduce a physical  length $\ell_{ph}$ between lattice points $n_1$ and $n_2$ 
and a physical (renormalized) mass $m_R$ by  
\beq\label{am4}
\ell_{ph}(n_1,n_2)  = a |n_1 -n_2| , \qquad m_R  = \frac{1}{a \xi}.
\eeq
This ensures that the exponential decay of the continuum, renormalized  two-point 
function is defined by $m_R$ since we have 
\beq\label{am5}
\e^{- m_{ph} \ell_{ph}(n_1,n_2)} = \e^{-|n_1-n_2|/\xi} 
\eeq
and for fixed $m_{R}$ \rf{am4} shows that $\xi \to \infty$ leads to $a \to 0$, i.e.\
a removal of the UV cut-off and the definition of a continuum quantum 
field theory with renormalized coupling constants $m_R$ and $\kp_R > 0$.

The above discussion assumes that the lattice is infinite since we are discussing 
limit where $\xi \to \infty$. In actual numerical lattice Monte Carlo simulations we 
are forced to have a finite lattice consisting of $N_4 = L^4$ lattice points. The correlation
length can then not be larger than $L = N_4^{1/4}$. However, assuming that 
$N_4$ is sufficiently large we have, according to  \rf{am2}: 
\beq\label{am3}
|\kp_0(N_4^{1/4})-  \kp_0^{uv}| =  c(\kp_R) N_4^{\beta_0'(\kp_0^{uv})/4},
\eeq
which is a so-called finite size scaling relation that we will use also in the case of 
lattice gravity. If we are in a regime of coupling constant space where finite size 
scaling is valid we could have replaced the $\xi^{-1}$ axis with a $N_4^{-1/4}$ axis,
or even a $N_4^{-1}$ axis. The qualitative features of Fig.\ \ref{fig-appendix}  
would be unchanged.
This is precisely what we will do in the case of lattice gravity, as will be explained below.

\section{Lattice quantum gravity: CDT}

\begin{figure}[t]
\vspace{-1mm}
\centerline{{\scalebox{0.6}{\includegraphics{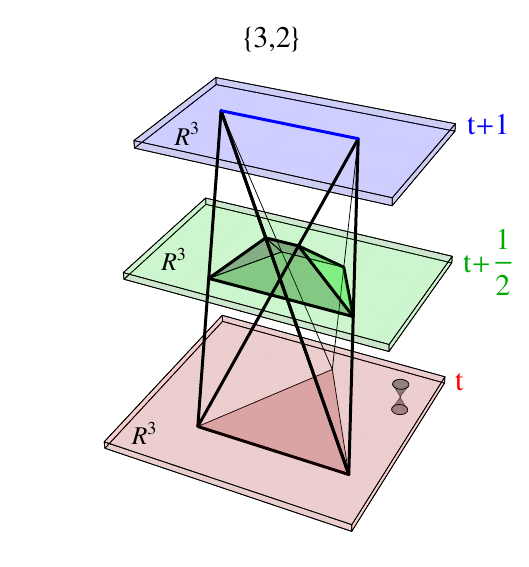}}} {\scalebox{0.6}{\includegraphics{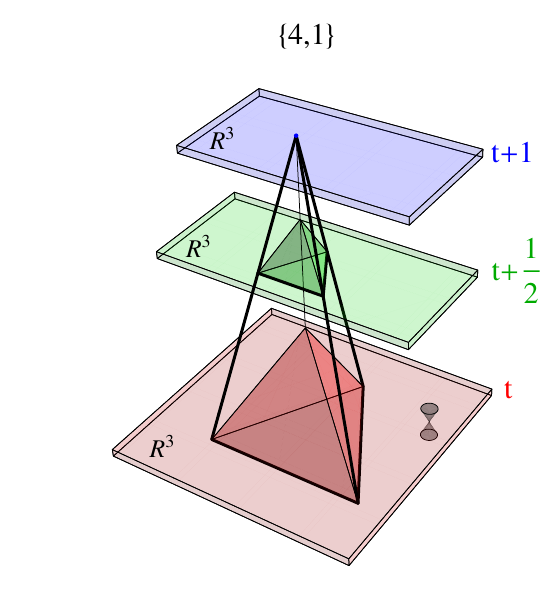}}}}
\caption{ \small The build-up of a CDT triangulation between the time-slab at $t$
and at $t+1$. Shown is a so-called (3,2) four-simplex and a $(4,1)$ four-simplex. }
\label{fig-am2}
\end{figure}

Four-dimensional  Dynamical Triangulations (DT)  and four-dimensional 
Causal Dynamical Triangulations (CDT) provide  
lattice regularizations of 4d quantum gravity (see \cite{review1,review2,reviewj,reviewjl} 
for reviews). For an ordinary lattice field theory, such as the $\phi^4$ theory discussed 
above, the lattice is fixed and the dynamics comes from the fields $\phi(n)$ 
living on the lattice points $n$. In DT and CDT  the dynamics comes from summing over different lattices. One considers 4d piecewise linear manifolds of a fixed topology, defined 
by gluing together identical building blocks of four-simplices with link length $a$, the 
$a$ acting as a UV cut-off as in the case of the $\phi^4$ lattice theory. Viewing the 
four-simplices as flat in the interior, a unique 
geometry is associated to  each such piecewise linear manifold since geodesic 
distances between two points are well defined. At the same time the gluing will 
define a four-dimensional triangulation, a lattice. The path integral 
of quantum gravity involves the integration over all geometries and it will 
now be represented as a sum over all such triangulations or lattices of the given 
topology. The action \rf{f1} used in the case of the $\phi^4$ lattice theory represents 
the simplest discretized version of the continuum $\phi^4$ action. For piecewise 
linear manifolds there exists a beautiful geometric discretization of the 
Einstein-Hilbert action, where the curvature in the four-dimensional case lives 
on the triangles of the four-dimensional triangulation, the so-called 
Regge action \cite{regge}. In the case where the 
triangulation is constructed by gluing together identical building blocks the Regge
action becomes exceedingly simple since it will just depend linearly on 
the total number of four-simplices  and the total number of triangles. 
We will use this simple action in
the definition of the path integral. The final ingredient entering in CDT is that 
we assume that geometries have a proper time foliation that we implement 
in the following way.  Let time be discretized. At each  time $t_i$ we have a spatial 
slice $\Sigma(t_i)$ with a fixed spatial topology. 
Here we consider the simplest case where the 
spatial topology is that of the three-sphere $S^3$. 
 We triangulate each $\Sigma(t_i)$ by gluing together tetrahedra to form an 
triangulation with the topology of $S^3$. 
We then fill out the slab  between $\Sigma(t_i)$ and $\Sigma(t_{i+1})$
by four-dimensional simplices, glued together in such a way that the topology of the 
slab is $S^3 \times [0,1]$. These four-dimensional triangulations can share a tetrahedron,
a triangle,  a link or a vertex with the three-dimensional triangulation of $\Sigma(t_i)$,
and they will then share a vertex, a link, a triangle or a tetrahedron with the three-dimensional triangulation of $\Sigma(t_{i+1})$, respectively. 
The construction is shown 
in Fig.\ \ref{fig-am2}. 

In the path integral we then sum over
all possible 3d triangulations of the spatial slices $\Sigma(t_i)$s and all possible 4d triangulations that fill out the slabs. 
Finally, each such triangulation $T_L$ is associated with 
a weight $e^{i S_{\rm regge}[T_L]}$, where $S_{\rm regge}[T_L]$ is the Regge action associated with $T_L$. The continuum path integral is then replaced by the following sum
\beq\label{am6}
Z(G,\Lam) = \int \cD [g] \; \e^{iS_{\rm eh}[g; G,\Lam)} \to 
Z_L(k_2,k_4) = \sum_{T_L} \e^{i S_{\rm regge} [T_L;k_2,k_4]}
\eeq
where the continuum Einstein-Hilbert action refers to the gravitational and cosmological 
coupling constants $G$ and $\Lam$, while the Regge action refers to the dimensionless
lattice analogues $k_2$ and $k_4$.

A special property of the CDT setup is that for each Lorentzian triangulation $T_L$ we 
can perform a  rotation to an Euclidean triangulation $T_E$,
simply by changing
the length assignment $l^2 = -a^2$ of the time-like links connecting $\Sigma(t_i)$ 
and $\Sigma(t_{i+1})$ to $l^2 = a^2$. Formally this is a rotation to imaginary time, i.e.\
Euclidean time. The Regge action will then change in the standard way:
\beq\label{am7}
i S_{\rm regge}[T_L] \to -S_{\rm regge} [T_E], \quad {\rm i.e.} \quad Z_E [k_2,k_4] =
 \sum_{T_E} \e^{- S_{\rm regge} [T_E;k_2,k_4]}.
\eeq
In the following we will always sum over this class of Euclidean triangulations and 
drop the subscript $E$. We then have a theory with Euclidean signature, like
in the $\phi^4$ case, but the class of geometries is smaller than the one 
provided by the full class 
of Euclidean triangulations since the triangulations $T_E$ that enter in \rf{am7} 
 still remember the time-slicing we imposed  on the triangulations\footnote{
 The four-dimensional DT lattice gravity formulation pre-dates the CDT formulation 
 \cite{aj,ajk}. In the DT theory one sums over the full class of 
 Euclidean triangulations. In this way one avoids introducing a time foliation. However,
 it is unclear how to relate the theory to a gravity theory with Lorentzian signature. Also,
 it was not clear how to obtain an interesting continuum limit of the DT lattice 
 theory, although this is still under investigation \cite{jack,jack1}. } $T_L$.

As stated above, the Regge action becomes very simple when one uses 
identical building blocks. In CDT we have, in a Wilsonian spirit, chosen to 
generalize the Regge action slightly by allowing different cosmological 
coupling constants associated with four-simplices of type (4,1) and type (3,2)
shown in Fig.\ \ref{fig-am2}. The action then becomes
\beq\label{am8}
S[T] = -k_2 N_2(T) + k_{32} N_{32}(T) + k_{41} N_{41}(T)
\eeq
where $N_2(T)$ is the number of triangles in $T$, $N_{32}(T)$ the number of $(3,2)$ plus 
$(2,3)$ four-simplices and $N_{41}(T)$ the number of $(4,1)$ plus $(1,4)$ four-simples.
The total number of four-simplices in $T$ is $N_4(T) = N_{41}(T) + N_{32}(T)$. Using 
the so-called Dehn-Sommerville relations between the number of subsimples $N_i(T)$
of order $i$, where $N_0(T)$ is the number of vertices, we can write \rf{am8} as follows
(see \cite{review1} for details) 
\beq\label{am9}
S[T;k_0,\Del,k_4] = -(k_0+6 \Del) N_0(T) + k_4 N_4(T) + \Del N_{41}(T).
\eeq
This is the action we will use in the regularized path integral:
\beq\label{am10}
Z[k_0,\Del,k_4] = \sum_T \e^{-S[T;k_0,\Del,k_4]}
\eeq
Here $k_0$ is formally related to the $a^2/G$ via Regge calculus, 
$\Del$ affects the ratio between $(4,1)$ and $(3,2)$ four-simplices, 
while $k_4$ monitors $N_4$,  the number of four-simplices.   

\subsection*{Coupling constants and correlation lengths}

The coupling constant $k_4$ in \rf{am9} plays a special role. This is seen by writing 
\rf{am10} as 
\beq\label{am11}
Z[k_0,\Del,k_4] = \sum_{N_4} \e^{-k_4 N_4}   Z[k_0,\Del;N_4]
\eeq
where $Z[k_0,\Del;N_4]$ denotes the partition function for a fixed $N_4$. It grows 
exponentially with $N_4$ and we can write
\beq\label{am12}
Z[k_0,\Del,k_4] = \sum_{N_4} \e^{-(k_4-k_4^c(k_0,\Del)) N_4} F(k_0,\Del;N_4),
\eeq
where $F$ is subleading as a function of $N_4$. We cannot perform the sum 
analytically and the only way to study the partition function is via Monte Carlo 
simulations and in these studies we are interested in testing as large $N_4$ as 
possible. In principle, by changing  $k_4$ in the neighborhood of 
$k_4^c(k_0,\Del)$ we can monitor $N_4$. However, it is much more convenient
to fix $N_4$ in the computer simulations. Then $k_4$ will play no active role, and 
to compensate for this we perform independent computer simulations for different $N_4$.
 In reality we are then studying $F(k_0,\Del;N_4)$ where we can choose to view 
 $N_4$ as a ``coupling constant''. This seems a little weird from the point of view 
 of ordinary lattice field theory where $N_4$ is simply the volume of spacetime.
 However, as we discussed in the case of the lattice $\phi^4$ theory, the correlation length
 $\xi$ played a dominant role when we wanted to study the continuum limit, and first 
 we exchanged the bare mass for the correlation length, and next, when the volume 
 $N_4$ was finite we changed the maximal correlation length with $N_4^{1/4}$ and 
 studied finite size scaling in the limit $N_4 \to \infty$. So even in that case 
 one could (under the right circumstances) view $N_4$ as a coupling constant and we were interested in the limit where this coupling constant went to infinity. Here, in the 
 case of gravity, we are of course also interested in the limit where $N_4$ goes 
 to infinity, but the first obvious question is: how can  this limit, $N_4 \to \infty$, 
 be related to a divergent correlation length?
 
\begin{figure}[t]
\vspace{-1.5cm}
\centerline{ {\scalebox{0.2}{\includegraphics{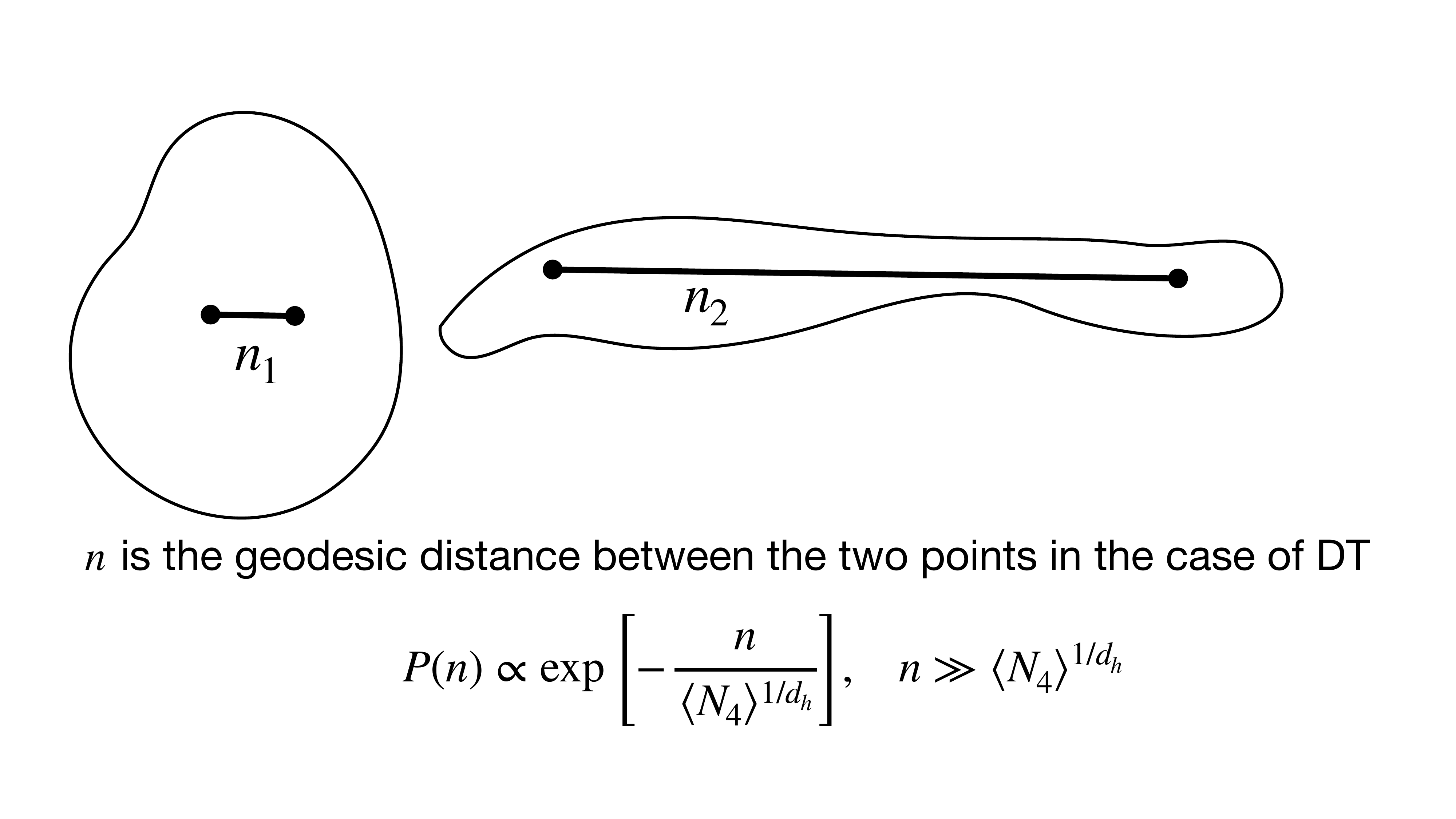}}}}
\vspace{-0.8cm}
\caption{\small Typical shape of a universe when $n$ 
is small and when $n$ is large.}
\label{fig-am3}
\end{figure}

 In fact, this question forces us to take a step back and ask the following question:
 how does one define the concept of a correlation length in a theory of quantum 
 gravity? In an ordinary QFT like the $\phi^4$ theory 
 one can define the correlation length as the exponential fall-off of the 
 two-point function $\la \phi(x) \phi(y) \ra$, as discussed above. It will be a function 
 of the (geodesic) distance between $x$ and $y$ (apart from lattice artifacts for small 
 lattice distances, if we consider a lattice version of the  $\phi^4$ theory). However, in 
 a theory of quantum gravity we are integrating over all metrics, and it is the metric 
 that determines the geodesic distance between two points $x$ and $y$. One way
 to define a (non-local) two-point function that {\it is} a function of a  geodesic distance 
 is the following:
 \bea\label{dis1}
G_\phi(D)& =& \int \cD [g]\!\int \cD \phi\;\e^{-S[g,\phi]}\times \\
&& \int\!\!\int  d^4x d^4y
 \sqrt{g(x)} \sqrt{g(y)}\; \phi(x) \phi(y)\; \del (D_g (x,y)\mi D),\non
\eea
In \rf{dis1} $D_g$ denotes the geodesic distance between points $x$ and $y$ measured
using the metrics $g$ used in the path integral. This formula has been shown to work 
well for Euclidean two-dimensional gravity coupled to conformal fields \cite{aa}. We 
are here going to apply it in the very simplest case where instead of fields $\phi(x)$
we just use the unit function $1(x) $. We then write
\beq\label{am13}
 G(D) = \int \cD [g]\;\e^{-S[g]} \int\!\!\int  d^4x d^4y
 \sqrt{g(x)} \sqrt{g(y)}\;  \del (D_g (x,y)\mi D).
 \eeq
Let $\la V_4 \ra$ denote the average four-volume of  our ensemble of geometries
we use in the path integral and let $d_H$ denote the Hausdorff dimension 
of the ensemble of geometries. Then, under quite general conditions one can show 
\cite{aj95} that the two-point function $G(D)$ falls off exponentially as 
\beq\label{am14}
G(D) = f(D) \; \e^{- c\, D/\la V_4\ra^{1/d_H}}, \qquad D \gg \la V_4 \ra^{1/d_H},
\eeq    
where $f(D)$ is subleading. The above expressions are readily translated
to the lattice formulation with $V_4$ replaced by $N_4$  and $D$ being 
replaced by the shortest graph distance $n$ in a triangulation and we write
\beq\label{am15}
G(n) = f(n) \; \e^{-c \, n / \la N_4 \ra^{1/d_H}}, \qquad n \gg \la N_4 \ra^{1/d_H}.
\eeq
The intuition behind the  fall-off is illustrated in Fig.\ \ref{fig-am3}:
the number of triangulations where two points are separated by a distance $n$ is 
a decreasing function of $n$.  The derivation in \cite{aw,ajw,book} for two dimensions
and in \cite{aj95} in four dimensions is for Euclidean 
quantum gravity. i.e.\ in the lattice version it is  for DT. In the case of CDT one has
to modify the proof because of the special role of the time direction. We will 
omit the details here. 

From eq.\ \rf{am15}  it is seem that $\la N_4 \ra^{1/d_H}$ plays the role of a 
correlation length for the two-point function $G(n)$.
Let $k_0,\Del$ be fixed. If $k_4$ can be chosen such that $\la N_4 \ra$ is 
 very large, we expect that most observables will have the same value 
 in the grand canonical ensemble with that chosen value of $k_4$ and in 
 the canonical ensemble where we fix $N_4 = \la N_4 \ra_{k_4}$ and thus 
 the interpretation of $N_4^{1/d_H}$ as a correlation length for the ensemble 
 of fluctuating geometries is a natural analogy to $N_4^{1/4}$ being the correlation 
 length for the ensemble of lattice $\phi(x)$ field configurations when 
 the dimensionless mass parameter is chosen such that correlation length 
 is equal to the linear size of the lattice. In the standard finite size scaling 
 scenario one chooses $N_4$ and then adjusts the bare mass parameter such that 
 the correlation length is equal to $N_4^{1/4}$ and the critical surface is 
 reached for $N_4 \to \infty$. In practical applications one 
 does not actually measure correlation length, but uses convenient scaling variables
 to observe finite size scaling, taking for granted that such scaling is only observed
 when the correlation length is comparable to $N_4^{1/4}$. In our CDT case 
 we will use the same philosophy: if we observe finite size scaling for some 
 observables, when comparing measurements for systems with different $N_4$,
 we will take it as a sign that $N_4^{1/d_H}$ can be used as the 
 correlation length and that the critical surface is reached when $N_4 \to \infty$.
  What is different in our  case 
 is that {\bf (1)} we cannot separate the correlation length from the (average) size 
 of the system and {\bf (2)}  the existence of the two-point function $G(n)$ with 
 a divergent correlation length does not imply that we have propagating degrees
 of freedom associated with this two-point function.

\subsection*{The CDT phase diagram}

\begin{figure}[t]
\centerline{\scalebox{0.2}{\rotatebox{0}{\includegraphics{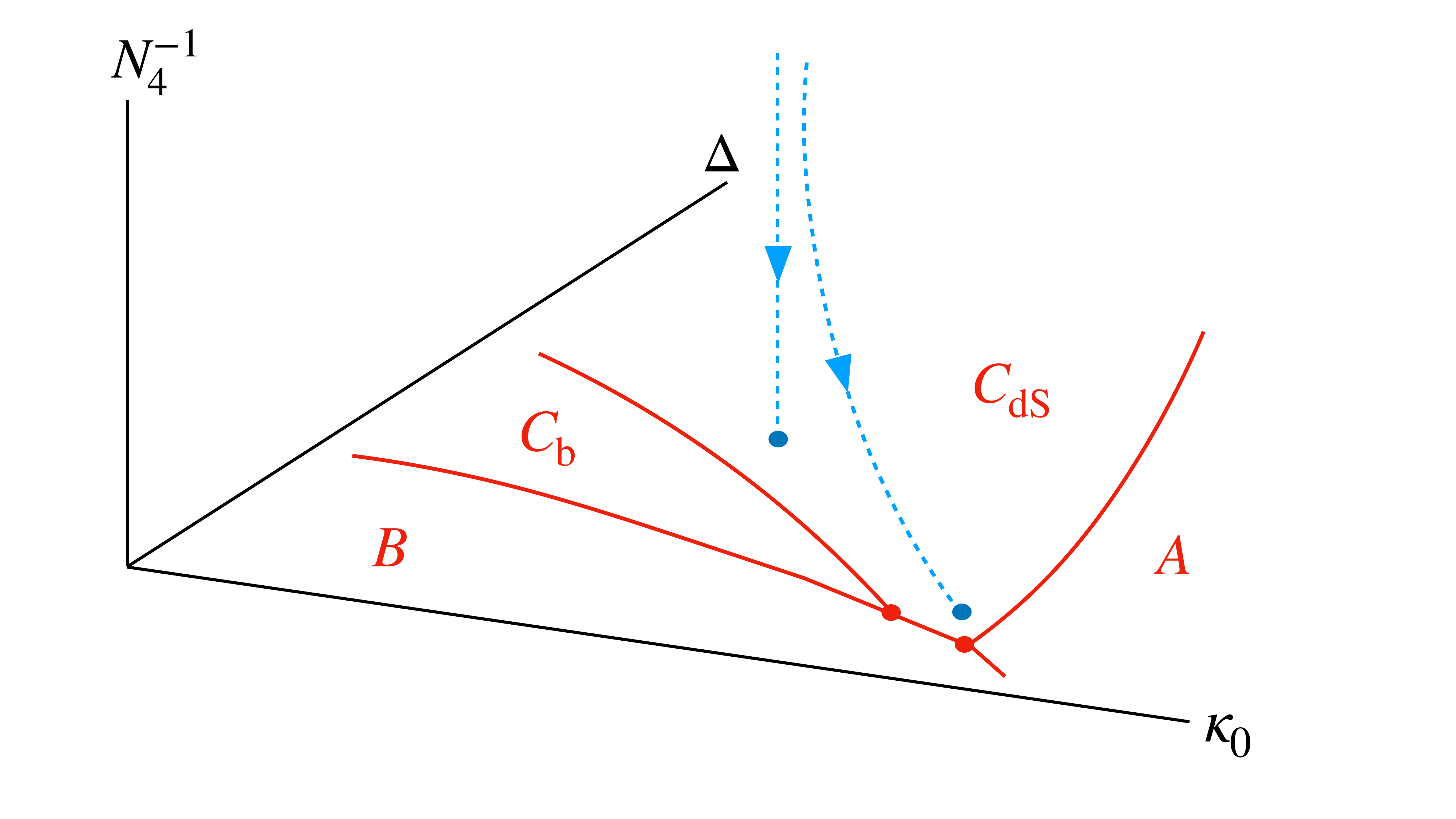}}}}
\vspace{-5mm}
\caption{{\small The CDT phase diagram where $N_4^{-1}$ is 
also included. Criticality can only occur when 
$N_4^{-1}= 0$. The straight vertical line corresponds
to keeping the bare lattice coupling constants $\kp_0,\Del$ fixed, while the other line is illustrating the flow when the renormalized coupling constants are fixed and one has to change the lattice coupling constants when approaching the critical surface.}}
\label{fig2-appendix}
\end{figure}

\begin{figure}[t]
\vspace{-1cm}
\centerline{\scalebox{0.8}{\rotatebox{0}{\includegraphics{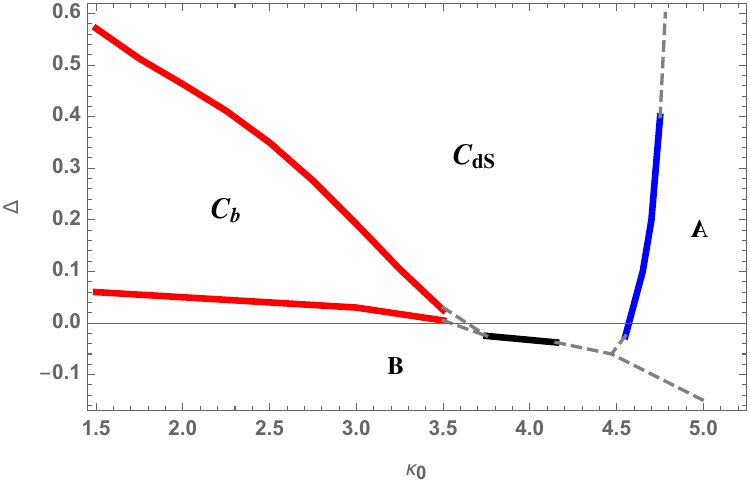}}}}
\caption{The CDT phase diagram. In phase $A$ different time slices seem not to couple.
In phase  $B$ the time extension of the universe is only one time-slice.
In phase $C_b$ the time extension of universe is larger, but it does not 
scale when $N_4$ is increased. Only phase  $C_{\rm dS}$ 
seems to represent a four-dimensional universe.}
\label{fig1}
\end{figure}

The Monte Carlo simulations using \rf{am10} reveal that there are a number 
of different phases in CDT, called $A$, $B$, $C_b$ and $C_{\rm dS}$ \cite{critical-lines}. 
We have coupling constants $k_0,\Del$ and we have 
$N_4$. The corresponding three-dimensional phase diagram is shown in 
Fig.\ \ref{fig2-appendix}. It should be compared
to the $\phi^4$ phase diagram  shown in Fig.\ \ref{fig-appendix}.
Only in the so-called de Sitter phase $C_{\rm dS}$ do we 
observe finite size scaling when $N_4 \to \infty$. Thus, only this phase will 
have our interest. We view the other phases as lattice artifacts. In Fig.\ \ref{fig1}
we show the surface corresponding (approximately) to $N_4 = \infty$.
The only part of this surface that we view as a critical surface is the part 
corresponding to phase $C_{\rm dS}$.

Let us now discuss how we observe finite size scaling in phase $C_{\rm dS}$ \cite{agjl}. 
 In the Monte Carlo 
simulations we have direct access to the three-volume $N_3(i)$, the number 
of three-simplices  at time-slice $i$. 
For a fixed $N_4$ we can now measure 
$\la N_3(i)\ra$ and $\la N_3(i_1) N_3(i_2)\ra$. For $N_3(i)$ we observe
for fixed $k_0,\Del$ that
\begin{figure}[t]
\centerline{{\scalebox{0.55}{\includegraphics{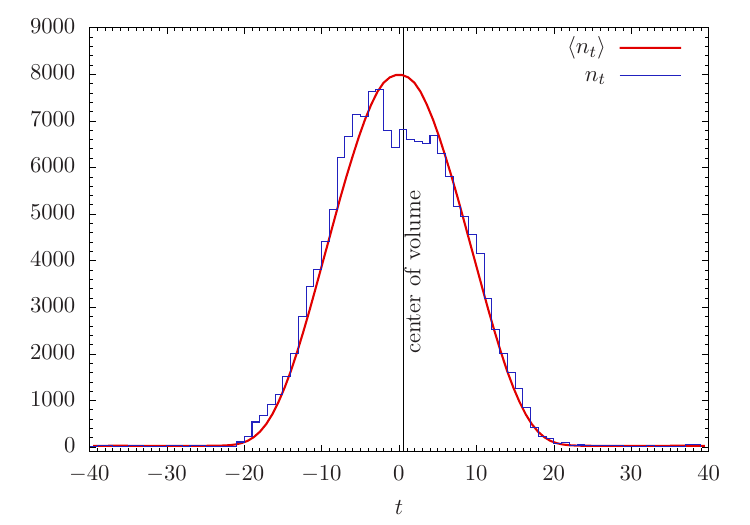}}} 
{\scalebox{0.55}{\includegraphics{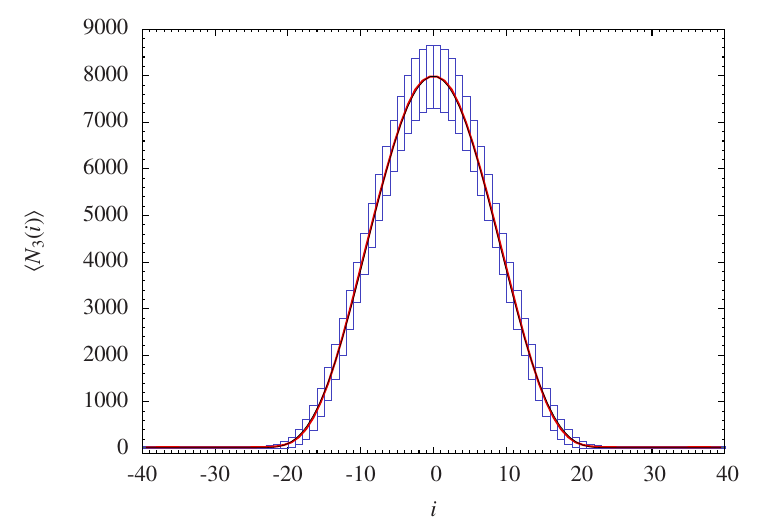}}}}
\caption{\small The left panel: 
A plot of a single $N_3(i)$ configuration for $N_4=362000$,  as well as the 
average of many configurations. The right panel: the plot of $\la N_3(i)\ra$ together 
with magnitudes   $ \sqrt{ \la \Del N_3(i) \Del N_3(i)\ra} $ of the fluctuations of $N_3(i)$.}
\label{fig-am4}
\end{figure}
\beq\label{am16}
\la N_3(i) \ra_{N_4} \propto N_4  \frac{1}{ \om N_4^{1/4}} 
\cos^3\left( \frac{i}{\om N_4^{1/4}}\right),
\eeq
see Fig.\ \ref{fig-am4}.  $\om$ depends on $k_0$ and $\Del$, but is independent 
of $N_4$ for $N_4$ sufficiently large. 
Eq.\ \rf{am16} shows finite size scaling with 
a Hausdorff dimension $d_H =4$. 
If we  introduce  {\it scaling variables} 
\beq\label{j14}
s_i = \frac{i}{N_4^{1/4}}, \quad n_3(s_i) = \frac{N_3(i)}{N_4^{3/4}},
\eeq 
we can write 
\beq\label{j12}
\la n_3(s) \ra = \frac{3}{4\om } \, \cos^3 \Big( \frac{ s}{\om}\Big)
\eeq
Similarly, the correlations behave like 
\beq\label{am17}
\la \Del N_3(i_1) \Del N_3 (i_2) \ra  = 
\G N_4 F\left( \frac{i_1}{\om N_4^{1/4}},\frac{i_2}{\om N_4^{1/4}}\right), \quad 
\Del N_3(i) = N_3(i) - \la N_3(i) \ra.
\eeq 
Expressed  in scaled variables we have 
\beq\label{am18}
\la \Del n_3(s_1) \Del n_3 (s_2) \ra  =   \frac{\G}{\sqrt{N_4} } \;
F\left( \frac{s_1}{\om},\frac{s_2}{\om}\right).
\eeq
Eqs.\ \rf{j12} and \rf{am18} are very well described by the following effective action:
\beq\label{j15}
S_{\rm eff}[k_0,\Del] = \frac{1}{\G} \sum_i  \left(
\frac{\big(N_3({i+1}) -N_3(i)\big)^2}{ N_3({i})\big)} + \del \, N_3^{1/3}(i) \right).
\eeq
or,  expressed in scaling variables (with $ds_i = 1/N_4^{1/4}$)
\beq\label{j13}
S_{\rm eff}[k_0,\Del] = \frac{\sqrt{N_4}}{\G} \int_{-\pi \om/2}^{\pi \om/2} ds \; \left(
\frac{(\dot{n}_3^2(s)}{ (n_3(s)} + \del \, n_3^{1/3}(s) \right), \quad
\int_{-\pi \om/2}^{\pi \om/2}  ds \, n_3(s) = 1.
\eeq  
The solution to the ``classical'' eom associated with $S_{\rm eff}$ is 
precisely \rf{j12} provided $\del$ and $\om$ are related as follows
\beq\label{am19}
\frac{\del}{\del_0} = \left(\frac{\om_0}{\om}\right)^{8/3}, \qquad 
\del_0 = 9 (2\pi^2)^{2/3}, \quad 
\om_0 = \frac{3}{\sqrt{2}} \frac{1}{\del_0^{3/8}}.
\eeq
If $\del = \del_0$ \rf{j12} represents a ``round'' $S^4$ sphere with four-volume 1.
We will denote \rf{j12} $n_3^{cl}(s)$ and the data are then well described
by $n_3^{cl}(s)$ and Gaussian fluctuations around $n_3^{cl}(s)$.

In the computer simulations producing these results we have kept $k_0$ and $\Del$ 
fixed and varied $N_4$, that is, we have followed straight blue path shown in 
Fig.\ \ref{fig2-appendix}. The effective action describing our data close to 
the surface $N_4 = \infty$ contains two effective coupling constants $\G$ and $\del$.
For $k_0,\Del$ in the interior of phase $C_{\rm dS}$, $\G$ and $\del$ will depend
on $k_0,\Del$, but will be independent of $N_4$ for $N_4$ sufficiently large. However, 
how large $N_4$ has to be before $\G(k_0,\Del,N_4)$ and $\del(k_0,\Del,N_4)$
 becomes independent of $N_4$ will depend on $k_0$ and $\Del$.
We will now compare these lattice gravity results to the simplest FRG results.

\section{FRG}

\begin{figure}[t]
\centerline{\scalebox{0.2}{\rotatebox{0}{\includegraphics{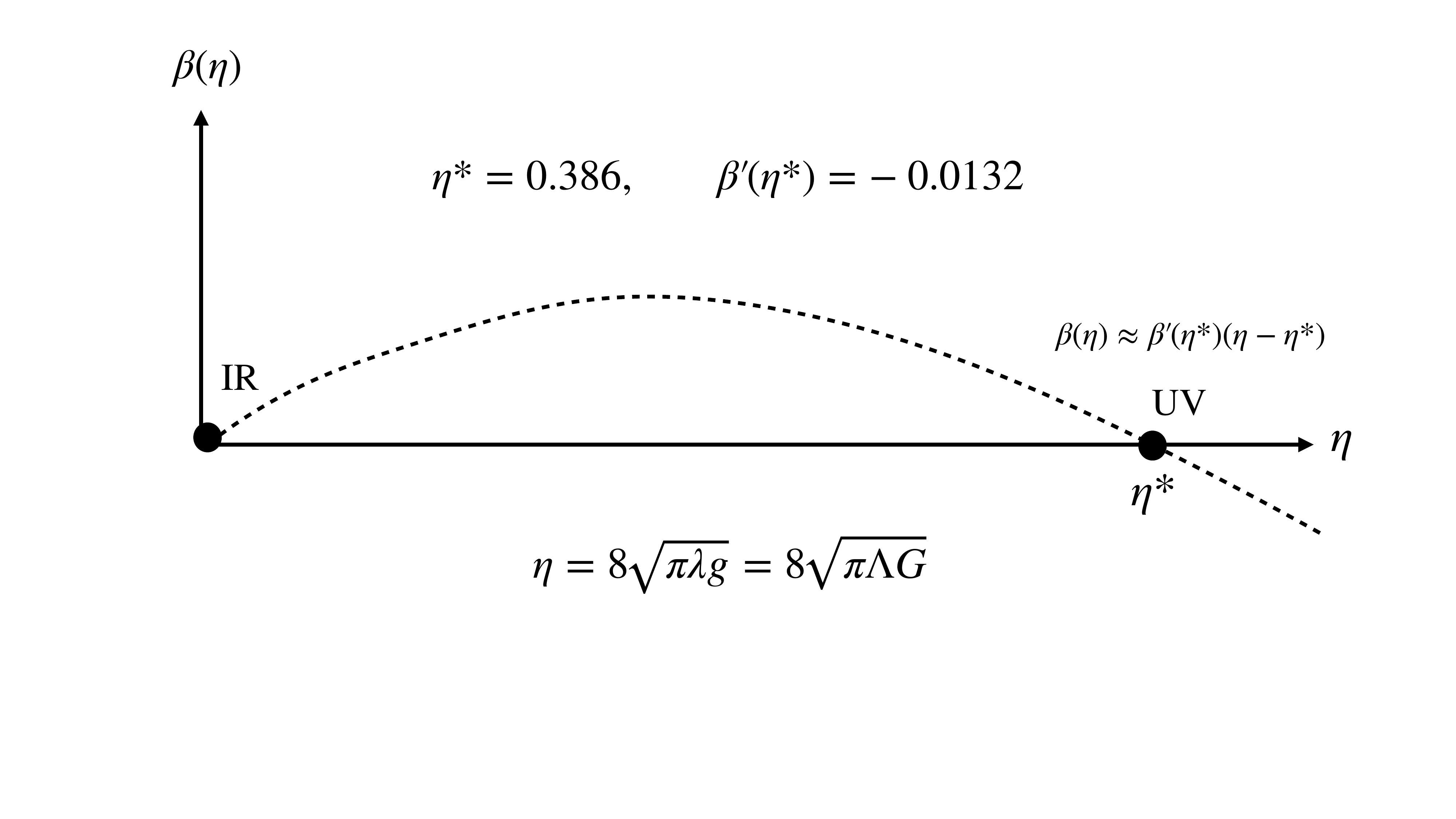}}}}
\vspace{-2cm}
\caption{\small A qualitative picture of the $\beta$-function provided in \cite{kawai}.}
\label{fig-am5}
\end{figure}

In the FRG approach one attempts to calculate an effective action as 
a function of a scale $k$. In actual calculations one uses a trial action with 
adjustable coefficients and tries to determine their $k$ dependence. Their behavior 
in the IR is then obtained for $k \to 0$, while the behavior in the UV is revealed 
for $k \to \infty$. The simplest effective 
action  considered is the  Einstein-Hilbert action where the gravitational 
constant $G$ and the cosmological constant $\Lam$ are functions of 
the  scale  $k$ that enters in the FRG:
\beq\label{j1}
\G_k[g_{\mu\nu}] = \frac{1}{16\pi G_k} \int d^4 x \sqrt{g(x)} \, \Big( -R(x) + 2 \Lam_k\Big).
\eeq
In \rf{j1} $\G_k[g_{\mu\nu}] $ is written for Euclidean signature of spacetime.
In the seminal work of Reuter \cite{reuter} it  was found that there is a UV fixed 
point for the ``running'' coupling constants $G_k$ and $\Lam_k$. More elaborate
calculations have not changed this conclusion (see \cite{review3,review4,review5} for 
details). Since the 
scale  $k$ has the dimension of mass we can write
\beq\label{j2}
G_k := g_k/k^2, \quad g_k \to g_*, \qquad \Lam_k := \lam_k k^2, \quad \lam_k \to \lam_*,
\eeq  
where $g_k$ and $\lam_k$ are dimensionless coupling constants that approach 
their UV fixed point values $g_*$ and $\lam_*$ for $k \to \infty$ and one  might 
try to compare $g_k,\lam_k$  to suitable dimensionless lattice gravity coupling constants.
In particular we have for the dimensionless combination $G_k \Lam_k$:
\beq\label{j3}
G_k \Lam_k  = g_k \lam_k \to g_*\lam_*\quad {\rm for} \quad k \to \infty.
\eeq
It has been argued \cite{kawai} that the dimensionless combination $G\Lam$ is the only
relevant coupling in the truncation \rf{j1}, or even for a more general 
class of truncations \cite{kevin}, and both
in \cite{kevin} and \cite{kawai} a $\beta$-function for $\eta=\sqrt{G \Lam}$ is found. 
In \cite{kawai} it is even provided as an explicit rational function of $\eta$, shown in 
Fig.\ \ref{fig-am5}, 
that  should be compared to Fig.\ \ref{fig-2am}. Around the UV fixed point
they behave qualitatively in the same way. The only difference is that the $\beta$-function
shown in Fig.\ \ref{fig-2am} is for the bare lattice coupling constant $\kp_0$, while the
$\beta$-function shown in Fig.\ \ref{fig-am5} is for the continuum, renormalized 
coupling constant $\eta$. The FRG is an equation for continuum, renormalized 
fields and coupling constants. According to eq.\ \rf{fj6}  the  $\beta$-functions 
for the bare and the renormalized coupling constants agree 
qualitatively, and one can show that they are identical to lowest non-trivial order 
at a UV fixed point.

We now  treat \rf{j1} as a standard effective action\footnote{In the actual FRG calculations 
one  often makes the decomposition $g_{\mu\nu} = g_{\mu\nu}^B + h_{\mu\nu}$,
where $g_{\mu\nu}^B$ is a fixed background metric (i.e.\ a fixed de Sitter metric) 
that is fixed even when the scale $k$ is changing. From first principles 
the effective action can only depend on $g_{\mu\nu}$, not the arbitrary choice 
$g_{\mu\nu}^B$. Our treatment here is the most naive implementation of what is 
suggested in \cite{fr1,fr2}, namely that the background one should use for a 
given scale $k$ should be the one that satisfies the equations of motion at that scale.
In  \cite{fr2} it called the choice of self-consistent  
background geometries.}
and find the extremum  for 
$\G_k[g_{\mu \nu}]$. It  is the (Euclidean)  de Sitter universe with 
cosmological constant $\Lam_k$, i.e. 
a four-sphere, $S^4$, with radius $R_k = 3/\sqrt{\Lam_k}$. This  four-sphere
has the four-volume
\beq\label{am20}
 V_4(k) = \frac{8 \pi^2}{3} R_k^4 = \frac{8 \pi^2}{3} \frac{81}{\lam_k^2} \frac{1}{k^4}
 \to \frac{8 \pi^2}{3} \frac{81}{\lam_*^2} \frac{1}{k^4} \quad {\rm for} \quad k \to \infty.
\eeq
In order to compare the FRG effective action with the CDT effective action
we will further restrict the effective action to only include global fluctuations
where $V_4(k)$ is kept fixed rather then $\Lam_k$ and write the 
corresponding minisuperspace action using a proper time metric:
\beq\label{j5}
ds^2 = dt^2 + r^2(t) d \Omega_3^2, \qquad V_3(t) = r^3(t) \int d\Om_3 = 2 \pi^2 r^3(t).
\eeq
The effective action for $r(t)$, or more conveniently $V_3(t) $, is then 
\beq\label{j6}
S_{\rm eff} = -\frac{1}{24 \pi G_k} \int dt \,\Big( \frac{\dot{V}_3^2}{V_3} + \del_0 V_3^{1/3} \Big),
\qquad \int dt \;V_3(t) = V_4(k).
\eeq

One can now 
study fluctuations around this solution and compare to the fluctuations 
observed in CDT \footnote{In \cite{knorr-frank}
it is shown that when calculating fluctuations for ``global'' quantities 
like the three-volume, only constant modes contribute when space is compact. 
These modes are 
precisely the  modes used when calculating fluctuations in the minisuperspace 
approximation.}. Introducing dimensionless variables $v_3 = V_3/V_4^{3/4}$
and $s = t/V_4^{1/4}$ we can write 
\beq\label{j8} 
S_{\rm eff} = -\frac{1}{24\pi} \frac{\sqrt{V_4(k)}}{G_k} 
\int ds\;\Big(  \frac{ \dot{v}_3^2}{v_3} + \del_0 v_3^{1/3} \Big), \qquad \int ds\, v_3(s) = 1.
\eeq
Here $s$ and $v_3(s)$ will be of order $O(1)$ and the ``classical'' solution
to the eom,   $v_3^{cl}(s)$, is the four-sphere with volume 1. We note  
that  the fluctuations around $v_3^{cl}(s)$ will for a given $k$ 
be governed by the effective coupling constant
\beq\label{j9}
g_{\rm eff}^2(k) =\frac{ 24 \pi \,G_k}{ \sqrt{V_4(k)}} = \frac{4}{\sqrt{6}} \,\Lam_k G_k
\approx 1.63 \,  \lam_k g_k.
\eeq
In the FRG analysis $\lam_k g_k$ is an increasing  function of $k$, but  even 
at the UV fixed point it is not  large. Thus, somewhat surprising, simple 
Gaussian fluctuations around $v_3^{cl}(s)$ seems to be a good approximation
all the way to the UV fixed point. This might explain  the related observation
mentioned above for  CDT.

\section{Comparing CDT and FRG}

 We want to compare the lattice effective action \rf{j13} and the FRG effective
 action \rf{j8}\footnote{A first such comparison was done in \cite{cdt-rgf}. 
 However, at that time the so-called bifurcation
 phase $C_b$ had not been discovered. It was viewed as part of phase 
 $C_{\rm dS}$.}$^,$\footnote{The alert reader might have noticed a disturbing 
 sign difference between \rf{j13} and \rf{j8}. We will argue that it is a good thing.
 Our lattice theory provides a regularization of the path integral and is finite. On the 
 other hand, the effective action \rf{j8} is sick since the kinetic term has 
 a wrong sign. This is why Hartle and Hawking made a further 
 analytic continuation \cite{hh}. Using (Euclidean) conformal time, they 
 made an analytic continuation of the conformal factor such that 
 the kinetic term changed sign. Using proper time instead of 
 conformal time this analytic continuation leads precisely from 
 \rf{j8} to \rf{j13}. Thus the CDT version of the effective action 
 {\it is} the Hartle-Hawking effective action.}. Let us for the moment ignore that
 $\del \neq \del_0$. Then it is natural to identify
 \beq\label{j16}
 \frac{\sqrt{N_4}}{\G(\kp_0,\Del,N_4)} = \frac{\sqrt{V_4(k)}}{24 \pi G_k} \approx
  \frac{1}{1.63\, \lam_k g_k}.
 \eeq
 Recall the discussion for the $\phi^4$ theory. A renormalized 
 coupling constant $\kp_R$ could take values between $\kp^{ir}_R$ and $\kp^{uv}_R$.
 In the $\phi^4$ theory, these values were obtained from the bare 
 coupling constants as shown in Fig.\ \ref{fig-appendix}. However, they could 
 equally well be obtained by solving the renormalization group 
 equation using the $\beta_R(\kp_R)$. This $\beta$-function would 
 look more or less like the $\beta$-function shown in Fig.\ \ref{fig-2am}, just 
 with  $\kp_0$ replaced by $\kp_R$.
 The renormalized running coupling constant would then run between 
 $\kp_R^{ir}$ and $\kp_R^{uv}$ and any value in this range will 
 qualify as the renormalized coupling constant, defined from 
 the bare lattice coupling constants when the continuum limit
 of the lattice theory is defined approaching the lattice UV fixed point.
 This is the way we will view \rf{j16}: the rhs is a renormalized coupling 
 constant and the lhs expresses how it is defined in terms of lattice 
 coupling constants, i.e. in terms of $k_0$, $\Del$ and the lattice 
 correlation length $N_4^{1/4}$ (for notational simplicity we will use $N_4$ instead of 
 $N_4^{1/4}$ below). In this context the scale  $k$ that appears in
 the continuum renormalization group equation just becomes a parametrization
 that determines how $\eta_k$ ``runs'' between $\eta^{ir}$ and $\eta^{uv} = \eta_*$
 shown in   Fig.\ \ref{fig-am5}.
 
 Recalling again the discussion surrounding Fig.\ \ref{fig-appendix}, 
 we have two ways to approach the critical surface $N_4 = \infty$: {\bf (1)} we can 
 keep $k_0,\Del$ fixed. Then the renormalized coupling, i.e. $\lam_k g_k$ should 
 flow to an IR fixed point and {\bf (2)} we keep the renormalized coupling 
 $\lam_k g_k$ fixed while approaching the critical surface $N_4 = \infty$. 
 This is only possible if we also change the bare coupling constants 
 $k_0,\Del$, and {\it if} it is possible to take $N_4 \to \infty$ while 
 keeping   $\lam_k g_k$ fixed, the bare couplings $k_0,\Del$ should 
 flow to a UV fixed point. If it is not possible, then there is no UV 
 fixed point\footnote{Recall that this was actually the case in the $\phi^4$ theory in 
 four dimensions.}.

\subsection*{The IR limit}

Let us first study case {\bf (1)}: we keep the bare coupling constants $k_0,\Del$
fixed and located in the interior of phase $C_{\rm dS}$. As already mentioned, this 
implies that $\G(k_0,\Del,N_4)$ (and $\del(k_0,\Del,N_4)$) 
will be independent of $N_4$ for sufficiently large 
$N_4$. From \rf{j16} it follows that when we approach the critical surface 
$N_4 = \infty$ then $\lam_k g_k \to 0$. Thus $\lam_k g_k =0$ should be an 
IR fixed point. Does this agree with the FRG picture? If both $\lam_k$ and $g_k$
go to zero we precisely approach the so-called Gaussian fixed point of the 
renormalization group flow and in fact lowest order perturbation theory tells us 
 (e.g. see the linear approximation to eq.\ (74) in \cite{review3}):
 \beq\label{ju}
 g_k = g_{k_0} \frac{k^2}{k_0^2}, \quad
 \lam_k = \Big(\lam_{k_0} \mi \frac{g_{k_0}}{8\pi} \Big)\, \frac{k_0^2}{k^2} + 
  \frac{g_{k_0}}{8\pi} \,\frac{k^2}{k_0^2}, \quad k \approx k_0, ~~ g_{k_0},\lam_{k_0} \ll 1.
  \eeq
 When $k \to 0$ then  $\lam_k \to \infty$ unless $g_{k_0} = 8\pi \, \lam_{k_0}$, 
 in which case
 we start out precisely at the unique renormalization group 
 trajectory that leads to the Gaussian
 fixed point. Unless that is the case, naive lowest order   perturbation theory will 
 become invalid for $k \to 0$, since $\lam_k \to \infty$. However, it has been 
 argued \cite{kevin} that using a somewhat more general setup, 
  called Essential Quantum Einstein Gravity, instead of the simple 
 effective action \rf{j1}, one obtains
 the Gaussian fixed point as the end point of a whole class of renormalization group
 trajectories for $k \to 0$. This makes the Gaussian fixed point a natural IR fixed point.  
  
  In addition the $k \to 0$ limit has been studied using FRG for time-foliated spacetimes
 \cite{frank-w}. This setup  is closer to the  CDT approach 
 and a new IR fixed point was found with the property that a 
 whole set of renormalization group trajectories starting out at 
 the UV fixed point will converge to this fixed point  for $k \to 0$:
 \beq\label{ju3}
 (g_k,\lam_k) \to (0, \oh) \quad {\rm for} \quad k \to 0.
 \eeq
 More precisely it was found that 
 \beq\label{ju4}
 g_k \propto \frac{k^4}{\tilde{k}^4}, \quad \lam_k \mi \oh \propto \frac{k}{\tilde{k}}, \quad {\rm for} \quad 
 k \to 0,
 \eeq
 where $\tilde{k}$ is some fixed, small scale.
 This scenario is  also compatible with the CDT  limit for 
 $N_4 \to \infty$ and $k_0,\Del$ fixed.
 This IR fixed point is different from the Gaussian fixed point since the approach to 
 the Gaussian fixed point can be parametrized by a classical gravitational coupling constant 
 $g_k/k^2 = G_k \to G_0$, while $\Lam_k \to 0$ as $ k^2$. For the other 
 IR fixed point we have $G_k \propto k^2 \to 0$, and also  $\Lam_k \propto k^2 \to 0$. 
 
  Again, it is instructive to compare to the $\phi^4$ lattice theory. Keeping the bare 
  coupling $\kp_0$ fixed and increasing the correlation length $\xi$ to infinity (or, 
  in a finite size scaling setup, $N_4 \to \infty$), we end up at a critical line
  associated to the IR fixed point: the renormalized coupling constant flows to 
 its IR fixed point value when we approach the $\xi = \infty$ line. Similarly here,
 keeping the bare coupling constants $k_0,\Del$ fixed and in the interior of 
 the $C_{\rm dS}$ region, the renormalized $\Lam G$ flows to its IR or Gaussian 
 fixed point and the whole interior $C_{\rm dS}$ region is thus associated with this 
 IR or Gaussian fixed point.

 \subsection*{The ultraviolet limit}
 
We now turn to scenario {\bf (2)} and try to  localize a putative  lattice UV fixed point.
We thus keep the rhs of eq.\ \rf{j16} fixed and try to find paths 
$N_4 \to \big(k_0(N_4),\Del(N_4)\big)$ such the  lhs of \rf{j16} stays
fixed for $N_4 \to \infty$. From the behavior of $\G(k_0,\Del,N_4)$ discussed 
above such a path has to lead to the boundary of the $C_{\rm dS}$ phase region
since $\G(k_0,\Del,N_4)$ stays finite for any $k_0,\Del$ in the interior of the 
$C_{\rm dS}$ phase. More precisely we only see a substantial increase 
of $\G(k_0,\Del,N_4)$ when $k_0,\Del$ approaches the $A \mi C_{\rm dS}$  boundary,
see Fig.\ \ref{fig2-appendix} and Fig.\ \ref{fig1}. 
This is thus where a possible UV fixed point has to be located. However,
before we discuss this in more detail we have to deal with the fact that 
$\del \neq \del_0$ in eq.\ \rf{j13}, since $\del(k_0,\Del)$ also increases a lot 
when we get close to the $A \mi C_{\rm dS}$ boundary, 
i.e.\ according to \rf{am19} $\om(k_0,\Del)$
decreases, implying that the time-extension of the four-dimensional computer universe 
shrinks.

\subsubsection*{Dealing with $\pmb{\del \neq \del_0}$ }

The measured values of $\om$ in the lattice simulations are in general 
different from the value $\om_0$ dictated by GR. 
Since we explicitly break the symmetry between space and time in our lattice regularization, we also have the freedom to scale space-like links and time-like links differently in order to obtain continuum results compatible with the spacetime symmetry present in GR. Denote the length of the time-like links by $a_t$ and the length of the space-like links by $a_s \equiv a$. The continuum three-volume of  a spatial slice at time $t_i \equiv a_t i$, consisting of $N_3(t_i)$ tetrahedra will then 
be $V_3(t_i) \propto N_3(t_i) a^3$. 
Similarly, the continuum four-volume of $N_4$ four-simplices will be 
$V_4 \propto N_4 a_t a^3$. Strictly speaking the situation is somewhat 
more complicated for the four-simplices. 
We refer to \cite{review1} and \cite{newarticle} for details.
However, for notational simplicity we will simply write
\beq\label{jw2}
V_4 = N_4 a_t a^3, \quad V_3 =  N_3 a^3.
\eeq
Then eq.\  \rf{j15}, where  $a$ was chosen to be 1, can be rewritten as 
\bea\label{jk6}
S &=& \frac{1}{\G} \sum_i \left(\frac{( N_3(t_i\plu a_t) \mi N_3(t_i))^2}{N_3(t_i)} + 
\del \;N_3^{1/3}(t_i)\right)\quad ~~( t_i \equiv a_t  i) \\
&=& \frac{a_t}{a^3\G} \sum_i  a_t \left( 
\frac{( V_3(t_i+a_t) \mi V_3(t_i))^2/a_t^2}{V_3(t_i)} + 
\frac{a^2}{a_t^2} \,{\del}\; V_3^{1/3}(t_i)\right),
\label{jk7}\\
&\to& \frac{1}{ 24 \pi G} \int dt \left( \frac{\dot{V}_3^2}{V_3} + \tilde{\del }\; V_3^{1/3} \right),
\quad \tilde{\del} = \frac{a^2}{a_t^2} \,\del, ~~ 24\pi G = \frac{a^3}{a_t}\,\G, ~~
\hspace{0.8cm}
\label{jk8}
\eea 
and 
\beq\label{jk9}
\sum_i N_3(i) = N_4 \to \int dt \,V_3(t) = V_4, \qquad 
V_4 = a_t a^3 N_4.
\eeq
where $\tilde{\del}$ and  $\om$ are related as in \rf{am19}: $\tilde{\del}\, \om^{8/3} =  \del_0 \om_0^{8/3}$. If $\om \neq \om_0$  the lattice  configurations are the ``deformed'' spheres because the time extension $N_t a_t= \om N_4^{1/4} a_t$ does not match 
the spatial extention $N_3^{1/3} a$, when we write $N_4 = N_t N_3$ and $a_t=a$.
We can correct that by writing
\beq\label{jk10}
a_t = \left(\frac{\om_0}{\om}\right)^{4/3} a\, .
\eeq
From Eq.\ \rf{jk8} it then follows that $\tilde{\del} = \del_0$ for the round $S^4$ and thus this choice of $a_t$ leads to an action $S_{\rm eff}$ given in \rf{j13} that we can identify with the FRG effective action \rf{j8} for some value of the scale parameter $k$. 

So given computer data $N_4,\om,\G$ we can associate a corresponding continuum, round  $S^4$ with four-volume $V_4$ and gravitation constant $G$ via:
\beq\label{jh4}
(N_4,\om,\G)  \to (V_4(k),\om_0,G_k), 
\eeq
 where
 \beq\label{jk11}
 V_4(k) = \left( \frac{\om_0}{\om} \right)^{4/3} N_4 a^4, \quad 
 24 \pi G_k =  \left( \frac{\om}{\om_0} \right)^{4/3} \G \, a^2
 \eeq
 and in particular
 \beq\label{jh2a}
\frac{\sqrt{N_4}}{\G} = 
\frac{\om^2}{\om_0^2}\; \frac{\sqrt{V_4(k)}}{24\pi G_k} \qquad 
{\rm or} \qquad 
\frac{\om^2 \G(k_0,\Del,N_4)}{\om^2_0 \sqrt{N_4}} \simeq 1.63\, \lam_k g_k.
\eeq

From what is said above it is clear that the only chance to satisfy \rf{jh2a}
for $N_4 \to \infty$ is by approaching the $ A \mi C_{\rm dS}$ transition line
from the $C_{\rm dS}$ side. We will discuss this below.

\subsubsection*{Scaling at the  UV limit}

 As discussed in the numerical results section below, the observed dependence on $\Del$ is weak in the region of interest, and for notational simplicity we will omit most 
 references to $\Del$ in the following. In this way the critical surface 
 $N_4 = \infty$ becomes a critical line, precisely as was the case for the $\phi^4$
 theory. This line is then naturally associated with the IR fixed point of the FRG, 
 in the same way as the critical line for the $\phi^4$ theory was associated 
 with a IR fixed point of the renormalized coupling $\kp_R$. We want to 
 investigate if there should be a lattice UV fixed point on the critical line.
 
 Approaching a point $(k_0,N_4 = \infty)$ on the critical line, we have for 
 all $k_0$ different from such a  UV critical point 
  $k_0^\UV$ that 
  \beq\label{am22}
  \G(k_0,N_4) \to \G(k_0) < \infty, \quad  \quad {\rm for} \quad N_4 \to \infty.
    \eeq
  \beq\label{am22a}
   \om(k_0, N_4) \to \om(k_0), \quad 0< \om(k_0)  < \infty
  \quad {\rm for} \quad N_4 \to \infty.
  \eeq

\begin{figure}[t]
\vspace{-1cm}
\centerline{\scalebox{0.2}{\rotatebox{0}{\includegraphics{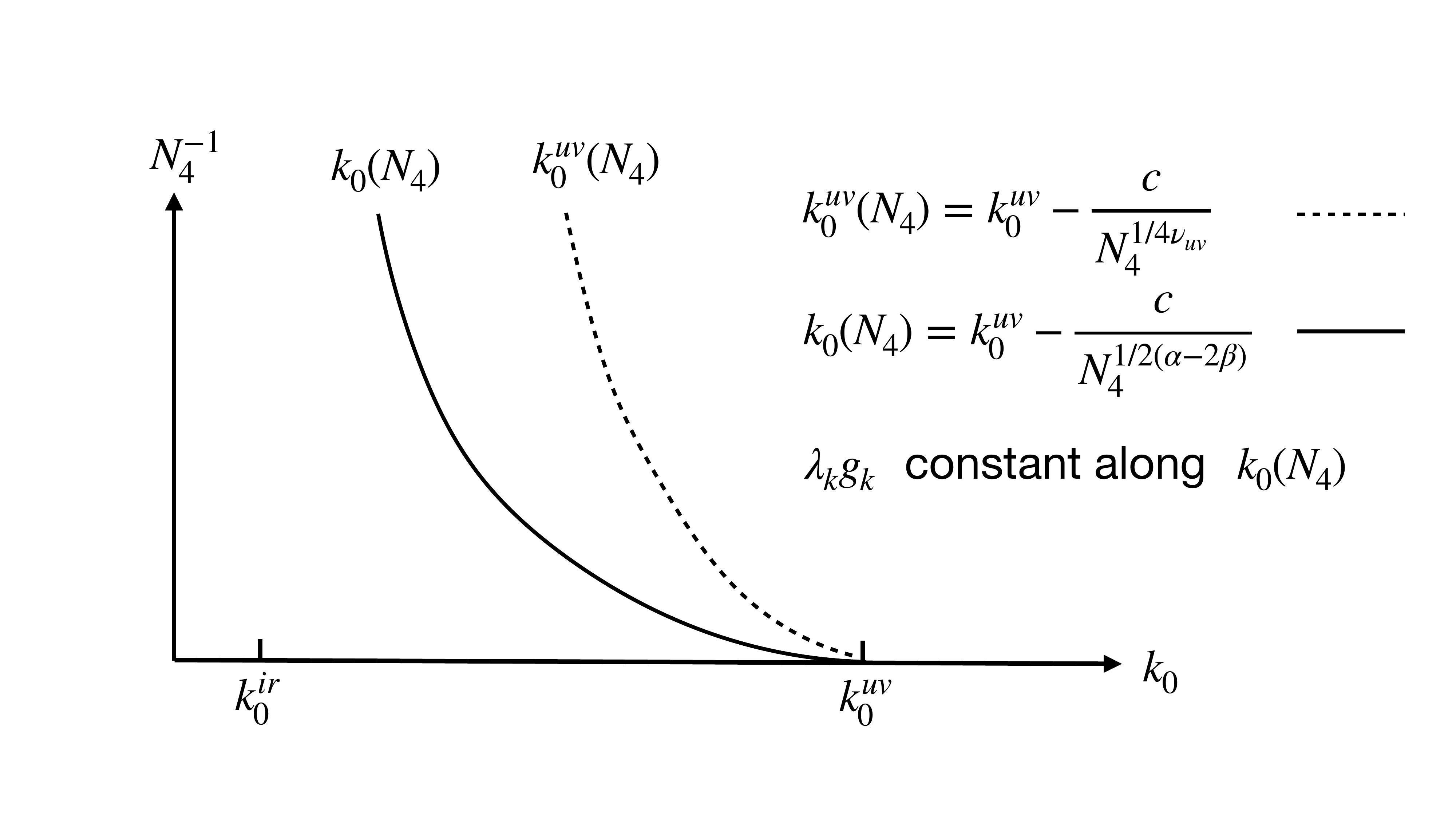}}}}
\vspace{-1cm}
\caption{{\small The tentative CDT phase diagram $(k_0,N_4^{-1})$ (with coupling constant $\Del$ ignored). Pseudo-criticallity appears along the dotted line $k_0^{\UV}(N_4)$ and the line $k_0(N_4)$, where $\lam_kg_k$ is constant is shown 
to the left of $k_0^{\UV}(N_4)$. The critical line is $N_4^{-1} = 0.$}}
\label{fig-am6}
\end{figure}

  The putative UV fixed point $k_0^{\UV}$ has to be located at the $A\mi C_{\rm dS}$ 
  transition line and we observe numerically that $\G(k_0) \to \infty$ and 
  $\om(k_0) \to 0$ for $k_0 \to k_0^{\UV}$. It is thus natural to assume that 
  close to $k_0^{\UV}$ we can have the following critical behavior
\beq\label{jo2}
\G(k_0) \propto \frac{1}{| k_0^{\UV} \mi k_0|^\alpha}, 
\quad \om(k_0) \propto | k_0^{\UV} \mi k_0|^\beta, \quad 
 \om^2(k_0)\G(k_0) \propto \frac{1}{| k_0^{\UV} \mi k_0|^{\alpha-2\beta}},
 \eeq
We further assume that for a finite $N_4$ there is a 
{\it pseudo-critical point} $k_0^{\UV} (N_4) < k_0^{\UV}$ where $\om^2(k_0,N_4)\G(k_0,N_4)$ has a maximum for fixed $N_4$, and that this pseudo-critical point approaches $k_0^{\UV}$ for $N_4 \to \infty$ as 
\beq\label{jo3}
k^{\UV}_0(N_4) = k_0^{\UV} -\frac{c}{N_4^{1/4\nu_{\UV}}} 
\qquad \Big( {\rm i.e.} 
\quad \xi \propto \frac{1}{|k_0^{\UV} \mi k_0^{\UV}(\xi)|^{\nu_{\UV}}} 
\Big).
\eeq
This implies that 
\beq\label{jo4}
\G(k_0^{\UV}(N_4) ) \propto N_4^{\alpha/4\nu_{\UV}}, 
\quad \om(k_0^{\UV}(N_4)) \propto N_4^{-\beta/4\nu_{\UV}}, 
 \eeq
 as well as 
 \beq\label{jo5}
 \om^2(k_0^{\UV}(N_4))\G(k_0^{\UV}(N_4)) \propto 
 N_4^{(\alpha-2\beta)/4\nu_{\UV}}.
\eeq
From Eq.\ \rf{jh2a} it follows that we have to have 
\beq\label{jo6}
\alpha - 2 \beta \geq 2 \nu_{\UV}
\eeq
and if that is the case the following path in the bare lattice coupling constant 
space will lead us to the putative UV fixed point while keeping $\lam_k g_k$ fixed:
\beq\label{jo7}
k_0(N_4) = k_0^{\UV} - \frac{c}{N_4^{1/2(\alpha - 2 \beta)}}
\eeq 
The situation is illustrated in Fig.\ \ref{fig-am6}.

\subsubsection*{The enigmatic relation between $\pmb{a}$ and $\pmb{k}$}

$k$ is a scale of dimension mass that appears in the FRG.  The dimensionless 
coupling constant $\lam_k g_k$ runs to the UV fixed point value $\lam_*g_*$ for 
$k \to \infty$. Similarly the inverse lattice spacing $a^{-1}$ is a UV cut-off scale that 
can be taken to infinity when the bare lattice coupling constants are approaching a
UV lattice fixed point: the ``continuum limit'' can be taken in such a way that 
the renormalized couplings are finite and non-trivial when $a \to 0$. It is thus natural
also to think about $k$ as a kind of UV cut-off \footnote{In the FRG community 
$k$ is often talked about as a UV cut-off, but it is also often emphasized that 
this should not be taken too literal, since  formally no UV cut-off is introduced 
explicitly when formulating the FRG.} such that $k \propto a^{-1}$ close to the UV 
fixed point. In our simple model we can address this. Recall that in our discussion 
so far $k$ only played a role as a parametrization of the renormalized 
coupling constant $\lam_k g_k$. Using \rf{jk11}, \rf{jo4} and $V_4(k) \propto \Lam_k^{-2}$
we find that 
\beq\label{jo9a}
a \propto \frac{1}{ \sqrt{\Lam_k}}\; N_4^{-\frac{1}{4}(1 + \frac{\beta}{3\nu_{\UV}})},
\quad {\rm i.e.} \quad a  \propto \frac{1}{ k}\, N_4^{-\frac{1}{4}(1 + \frac{\beta}{3\nu_{\UV}})} 
\quad {\rm for} \quad k \to \infty.
\eeq
Thus the lattice spacing $a$ scales to zero for a fixed value of $k$ (i.e.\ 
a fixed value of $\lam_k g_k$) when we approach the critical surface $N_4 = \infty$.   
It also follows from \rf{jo4} and \rf{jk10} that 
\beq\label{jo10} 
a_t \propto \frac{1}{\sqrt{\Lam_k}} \; N_4^{-\frac{1}{4}( 1- \frac{\beta}{\nu_{\UV}})}
\quad {\rm i.e.} \quad a_t  \propto \frac{1}{ k}\, N_4^{-\frac{1}{4}(1 - \frac{\beta}{\nu_{\UV}})} 
\quad {\rm for} \quad k \to \infty.
\eeq
This slower decrease of $a_t$ is a reflection of the fact that we, when approaching the $A\mi C_{\rm dS}$ transition line, have to rescale our lattice four-spheres that become 
increasingly ``contracted'' in the time direction,  
in order to match the round four-spheres of the FRG. Thus, under the assumptions 
\beq\label{am23}
\frac{\alpha -2\beta}{4 \nu^\UV} > \oh, \qquad \frac{\beta}{4 \nu^\UV} < \frac{1}{4}
\eeq
we can reach a UV fixed point where we can also take a the continuum limit 
$a,a_t \to 0$ for $N_4 \to \infty$. 

While \rf{jo9a} and \rf{jo10} tell us that $a \propto 1/k$, this is unfortunately just 
a dimensional relation. The real content of \rf{jo9a} is that for fixed $k$, i.e.\ for 
fixed $\lam_k g_k$, the lattice UV cut-off $a$ goes to zero when the 
correlation length $\xi$ (related to $N_4$) goes to infinity. In this sense 
it corresponds to the relation \rf{am4} for the $\phi^4$ lattice theory, $k$ playing the role
of $m_R$.

\section{Numerical results}

This Section will be rather short since the purpose of these Lectures 
has not been to discuss technical details of how to obtain the numerical 
results. For information about that we refer to \cite{newarticle} and references
therein. 

\begin{figure}[t]
    \centering
    \includegraphics[width=0.32\linewidth]{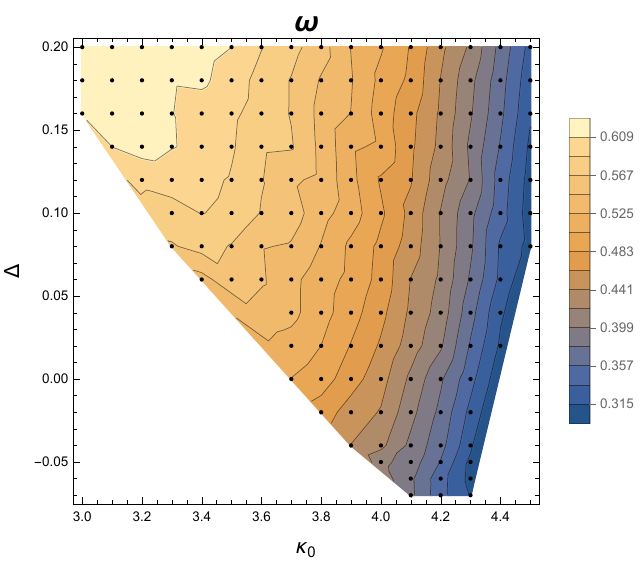}
    \includegraphics[width=0.32\linewidth]{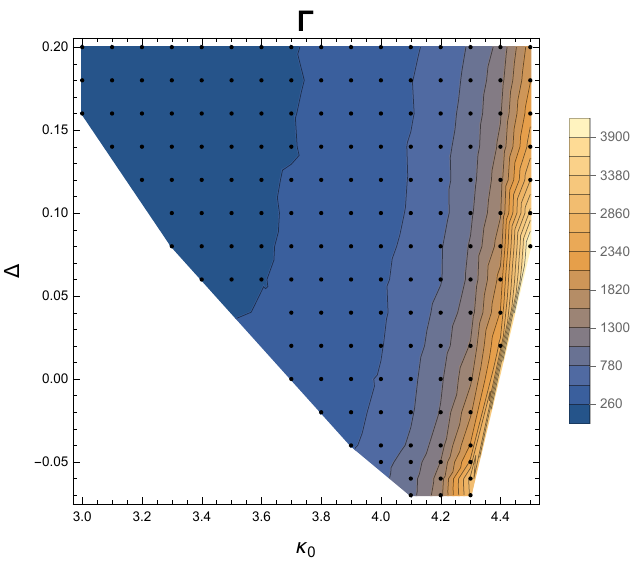}
     \includegraphics[width=0.32\linewidth]{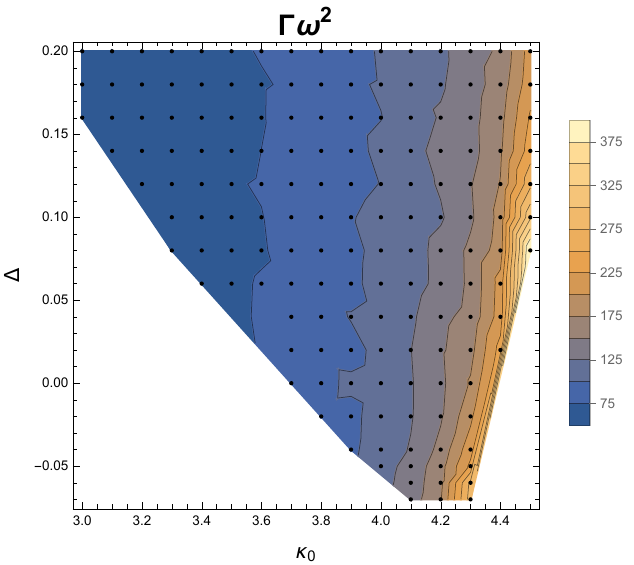}
    \caption{\small Contour plots of $\omega$ (left), $\Gamma$ (middle) and 
    $\om^2 \G$ (right) as  functions
     of the CDT coupling constants $k_0, \Delta$. Points where actual measurements were done are denoted as black dots in the plots.}
    \label{fig3}
\end{figure}

In Fig.\ \ref{fig3} we show the measurements of $\om(k_0,\Del)$, 
$\G(k_0,\Del)$ and $\om^2(k_0,\Del) \G (k_0,\Del)$ for a fixed $N_4$. One observes the 
increase of $\G$  and $\om^2 \G$ and the decrease of $\om$ when moving towards
the $A \mi C_{\rm dS}$ transition line. Also, the insensitivity to $\Del$ is seen.

\begin{figure}[t]
    \centering
    \includegraphics[width=0.32\linewidth]{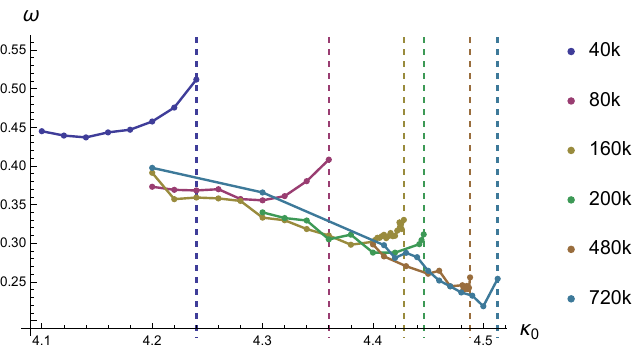}
    \includegraphics[width=0.32\linewidth]{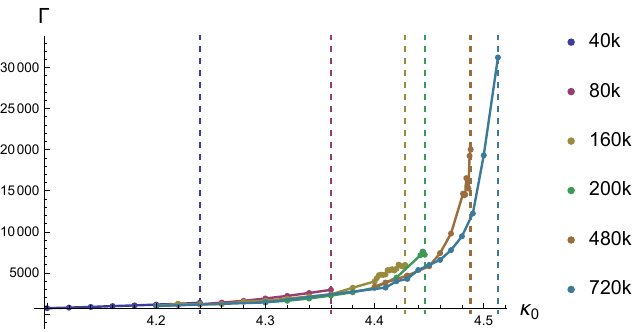}
    \includegraphics[width=0.32\linewidth]{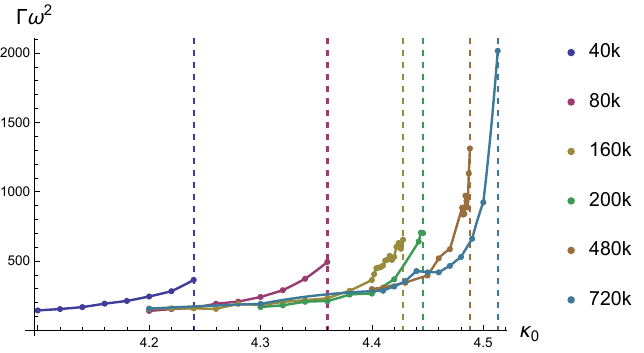}
    \caption{\small Dependence of $\omega$ (left), $\Gamma$ (middle) and $\Gamma \omega^2$ (right) on $k_0$ for fixed $\Delta=0$ measured for  the number 
    of (4,1)-simplices being $40\, 000, 80\, 000, 160\, 000, 200\, 000, 480\, 000, 720\, 000$ (denoted by different colors). Positions of $k_0$ closest to the pseudo-critical points $\kappa_0^{\UV}(N_4)$ are denoted by dashed lines.}
    \label{fig4}
\end{figure}

Fig.\ \ref{fig4} shows the measurements of the same observables at the pseudo-critical 
points $k_0^\UV(N_4)$. Close to  $k_0^\UV(N_4)$ the change of the observables
as a function of $k_0$ is fast, a fact that is not so visible in Fig.\ \ref{fig3}.

These measurements provide us with both $k_0^\UV(N_4)$,  
$\G(k_0^\UV(N_4))$, $\om(k_0^\UV(N_4))$ and 
$\om^2(k_0^\UV(N_4)) \G(k_0^\UV(N_4))$ and we can then determine the 
critical exponents defined in eqs.\ \rf{jo3}-\rf{jo5}. The determination of 
the exponents $\alpha/4\nu^\UV$,
$\beta/4 \nu^\UV$ and $(\alpha - 2\beta)/4\nu^\UV$ is shown in Fig.\ \ref{fig5} and 
the results  are:
 \begin{figure}[t]
    \centering
    \includegraphics[width=0.32\linewidth]{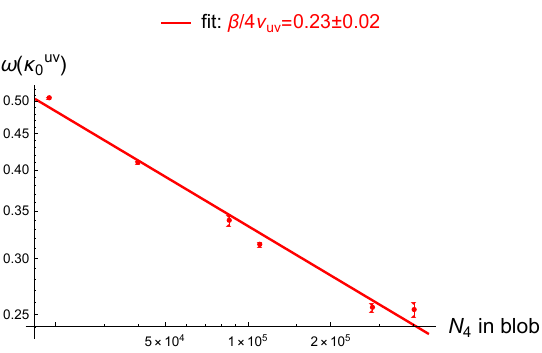}
    \includegraphics[width=0.32\linewidth]{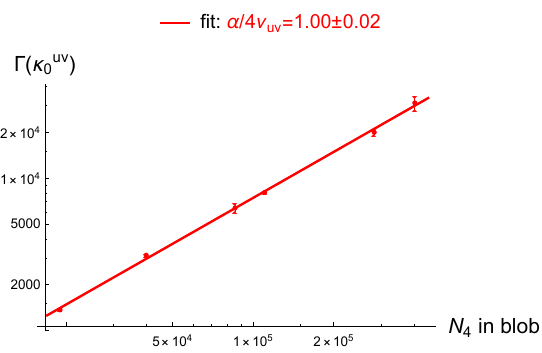}
    \includegraphics[width=0.32\linewidth]{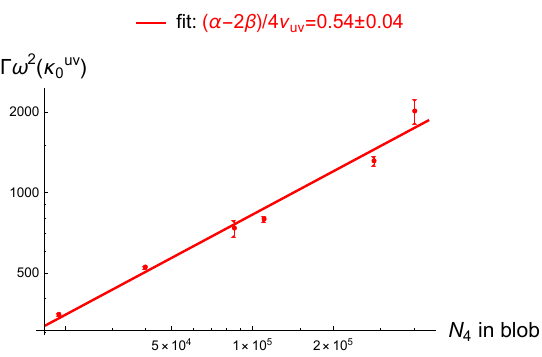}
    \caption{\small Critical scaling of $\omega$ (left), $\Gamma$ (middle) and $\Gamma \omega^2$ (right) measured closest to the pseudo-critical points $\kappa_0^{\UV}(N_4)$ (see Fig.~\ref{fig4}) for fixed $\Delta=0$. Fits of Eqs. \rf{jo4}-\rf{jo5} are depicted by solid lines. The figures show scaling as a function of the 
  $N_4$ volume contained in the $S^4$ ``blob'' (not all four-simplices are 
  contained in the $S^4$ part. Some can be in the so-called stalk, see Fig.\ 
  \ref{fig-am4}).}
    \label{fig5}
\end{figure}
 \beq\label{ceN4blob}
\frac{\beta}{4\nu_{\UV}} = 0.23 \pm 0.02, \quad \frac{\alpha}{4 \nu_{\UV}} 
= 1.00 \pm 0.02, \quad  \frac{\alpha-2\beta}{4 \nu_{\UV}} = 0.54\pm 0.04. 
\eeq
 What is striking about these results is that they are very close to the limit 
\rf{am23}. Thus we can say that the data allows for the existence of a UV fixed point,
but it cannot be used as a strong evidence for such a fixed point.

\section{Discussion}

We have tried to relate the simplest FRG flow to the CDT 
effective action for the scale factor of the universe. By using the analogy to
the $\phi^4$ lattice theory we argued that the $N_4 \to \infty$ limit 
of CDT, when in the $C_{\rm dS}$ phase, could be viewed as the critical 
surface associated with the Gaussian fixed point or an IR fixed point 
of the FRG theory. Again inspired by the $\phi^4$ lattice theory we then searched
for a CDT UV fixed point by studying the flow of the lattice coupling constants when 
the corresponding continuum coupling constants were kept fixed. Rather 
frustratingly the numerical accuracy is not yet good enough to decide if such 
a UV fixed point exists in CDT.

\subsection*{Acknowledgments}
It is a pleasure to thank Frank Saueressig,  Renata Ferrero, Martin Reuter and 
Kevin Falls for  enlightening discussions about FRG. DN is supported by the VIDI programme with project number VI.Vidi.193.048, which is financed by the Dutch Research Council (NWO). The research has been supported by a grant from the Priority Research Area (DigiWorld) under the Strategic Programme Excellence Initiative at Jagiellonian University.




\end{document}